\documentclass[9pt,twocolumn,twoside]{osajnl}
\usepackage{lineno}
\usepackage{float}
\usepackage{hyperref}
\usepackage{setspace}
\journal{ol} 

\setboolean{shortarticle}{true}
\title{Loading-effect-based 3-D microfabrication empowers on-chip Brillouin optomechanics}
\author[1]{Peng Lei}
\author[1]{Mingyu Xu}
\author[1]{Yunhui Bai}
\author[1]{Zhangyuan Chen}
\author[1,*]{Xiaopeng Xie}

\affil[1]{State Key Laboratory of Advanced Optical Communication Systems and Networks, School of Electronics, Peking University, Beijing 100871, China}
\affil[*]{Corresponding author: xiaopeng.xie@pku.edu.cn}

\begin{abstract}
The acousto-optic interaction known as stimulated Brillouin scattering (SBS) has emerged as fundamental principles for realizing crucial components and functionalities in integrated photonics. However, the main challenge of integrated Brillouin devices is how to effectively confine both optical and acoustic waves. Apart from that, the manufacturing processes for these devices need to be compatible with standard fabrication platforms, and streamlined to facilitate their large-scale integration. Here, we demonstrate a novel suspended nanowire structure that can tightly confine photons and phonons. Furthermore, tailored for this structure, we introduce a loading-effect-based three-dimensional microfabrication technique, compatible with complementary metal-oxide-semiconductor (CMOS) technology. This innovative technique allows for the fabrication of the entire structure using a single-step lithography exposure, significantly streamlining the fabrication process. Leveraging this structure and fabrication scheme, we have achieved a Brillouin gain coefficient of 1100 W$^{-1}$m$^{-1}$ on the silicon-on-insulator platform within a compact footprint. It can support a Brillouin net gain over 4.1 dB with modest pump powers. We believe that this structure can significantly advance the development of SBS on chip, unlocking new opportunities for the large-scale integration of Brillouin-based photonic devices.
\end{abstract}

\setboolean{displaycopyright}{true}
\begin{document}
\maketitle
Stimulated Brillouin scattering is a unique third-order nonlinear optical effect originating from optomechanical interactions \cite{Review1,Review2}. This effect stimulates acoustic waves across frequencies ranging from MHz to GHz and presents a distinctive narrowband gain resonance. SBS has found extensive applications in signal processing and microwave photonics, including millimeter-wave narrowband filters \cite{Filter2,IMP5}, narrow-linewidth lasers \cite{Laser4,IOSP3}, and distributed sensing \cite{Sensor1,Sensor3}. Moreover, leveraging its special acousto-optic coupling characteristics, Brillouin optomechanics has recently showcased substantial potential in the fields of optomechanical cooling \cite{cool}, non-reciprocity \cite{IOSP2}, and exceptional points \cite{Exc} within optical systems.

\begin{figure*}[ht]
\centering
\includegraphics[width=18cm]{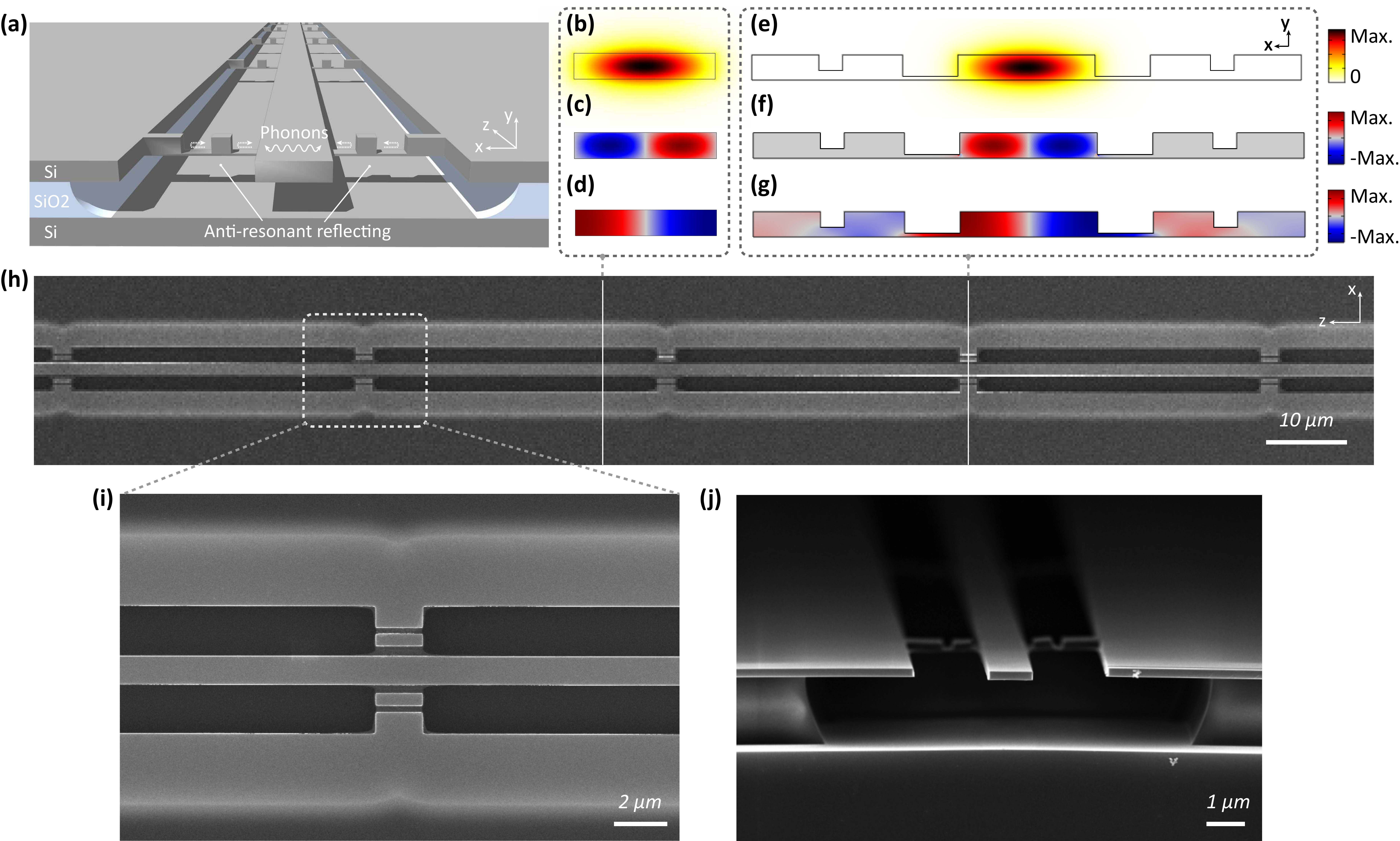}
\caption{(a) Schematic of the suspended nanowire. (b) and (e) are the electric fields of the optical modes. (c) and (f) are the x component of the electrostrictive force. (d) and (g) are the x component of the elastic displacement fields of acoustic modes. (b)-(d) depict the cross-sectional physical fields of the fully suspended nanowire, and (e)-(g) are the situation at the anti-resonant supporting structures. (h)-(i) Top-down scanning electron microscope (SEM) images of the fabricated device. (i) is the magnified view. (j) Cross-sectional SEM image of the fabricated device.}
\label{figure1}
\end{figure*}

These diverse applications typically demand flexible manipulation of the coupling strength between optical and acoustic modes. Benefiting from the advancements in integrated photonics, we can now engineer specialized waveguide structures flexibly, which enables the strength of Brillouin nonlinearity to far exceed that in optical fibers. In the past decade, experimental demonstrations have been conducted on various integrated photonic platforms, including chalcogenide glass \cite{Mat1}, silicon \cite{Mat3,Mat4}, silicon nitride \cite{Mat9}, and aluminum nitride \cite{Mat5}. Attributed to the compatibility of silicon waveguides with CMOS technology and their superior Brillouin gain coefficient (G$_B$) compared to other materials, silicon-on-insulator has emerged as a preferred Brillouin integrated platform. Due to the leakage of acoustic waves into the silica substrate, various suspension or isolation design strategies of silicon waveguides have been proposed in previous works \cite{Mat3,Mat4,PP2,Structure1}.

Nonetheless, the design strategy needs to consider the following factors. First, an ideal structure should be amenable to facile fabrication, CMOS-compatible, and capable of streamlining fabrication steps while improving the yield rate. Second, the structural design requires refinement based on actual fabrication and experimental results. Due to limitations in the mechanical quality factor (Q$_m$) caused by sidewall roughness and inhomogeneous broadening resulting from fabrication defects and the waveguide layout, the actual measured G$_B$ generally falls considerably short of the simulated estimates. Third, balancing the G$_B$ and optical losses is crucial for achieving Brillouin net gain. Although a reduction in the effective mode field area may yield a larger G$_B$,  it concurrently leads to a notable increase in optical scattering losses and nonlinear absorption losses. Meanwhile, the design of the supporting structure for the suspended waveguide must be meticulous, as optical waves passing through the supporting structure can result in mode-mismatch losses. However, most prior structures have struggled to address these issues comprehensively.

In our recent work, we introduced a suspended anti-resonant acoustic waveguide with superior confinement and manipulability of optical and acoustic modes \cite{SARAW}. Additionally, we developed a loading-effect-based three-dimensional microfabrication technique. It streamlined the entire waveguide fabrication process with only one lithography exposure and one plasma etching step. In this work, we further introduce a novel suspended nanowire waveguide, where the supporting structure adopts an anti-resonant design philosophy. Moreover, we have refined the loading-effect-based etching technique for this structure, facilitating remarkably straightforward design and fabrication processes. Leveraging this structure, we have achieved a forward Brillouin gain coefficient of 1100 W$^{-1}$m$^{-1}$ using a 5-cm-long waveguide with a compact footprint of merely 1200 $\mu m \times$ 1200 $\mu m$. With modest pump powers, it can support an on-off gain of 8.7 dB and a net gain of 4.1 dB, which represents the state-of-the-art level of on-chip Brillouin amplification in silicon waveguides.

\begin{figure}[ht]
\begin{center}
\centering
\includegraphics[width=8cm]{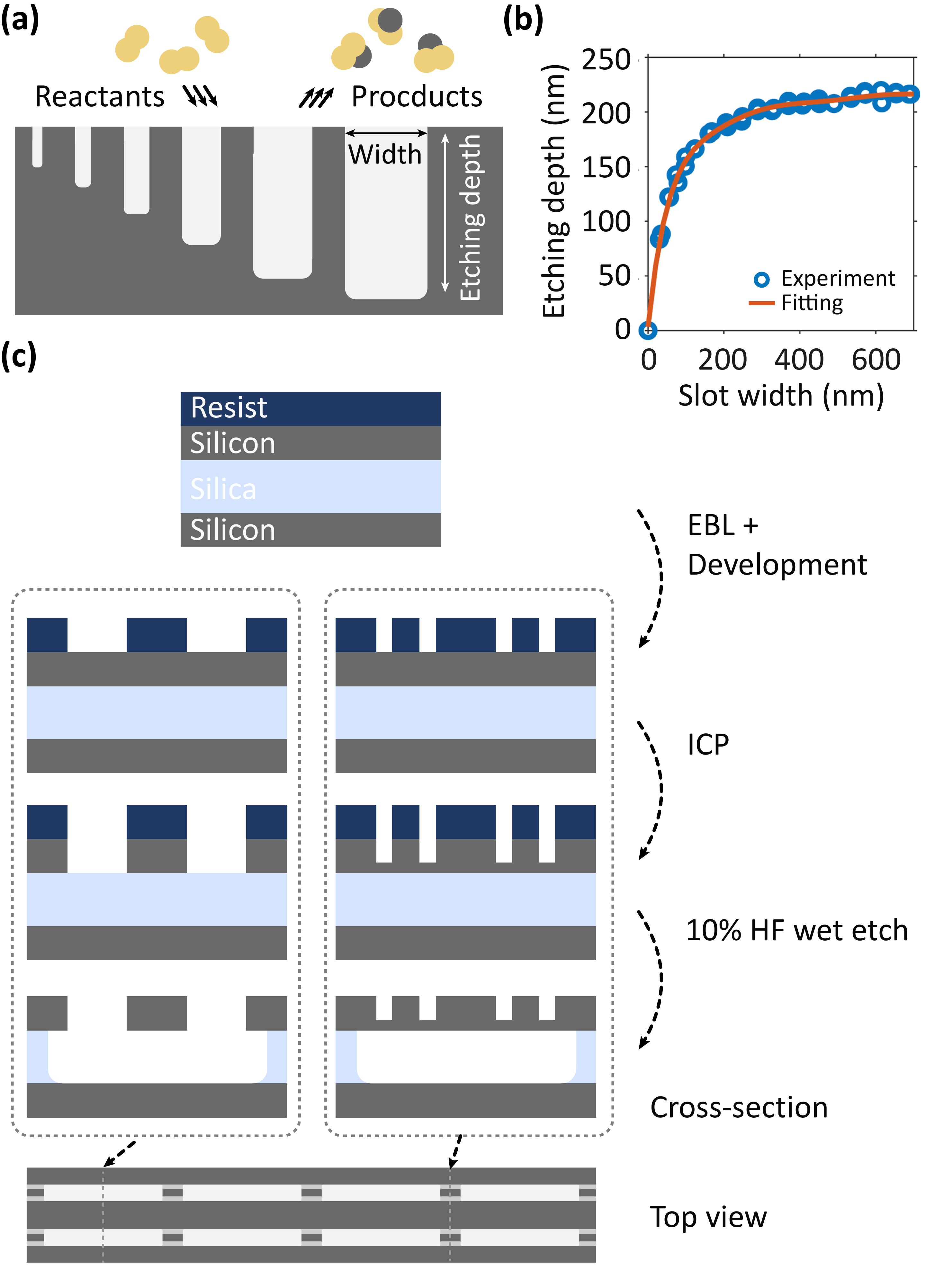}
\caption{(a) Diagram of the loading effect. (b) Experimental and fitting data of the loading effect. (c) Fabrication process. The dashed boxes on the left and right sides respectively illustrate the variation of the suspended nanowire and the anti-resonant supporting structure during the fabrication steps. EBL: electron beam lithography; ICP: inductively coupled plasma; HF: hydrofluoric acid solution.}
\label{figure2}
\end{center}
\end{figure}

The suspended nanowire is composed of a central rectangular waveguide and adjacent anti-resonant supporting arms, as illustrated in Fig. \ref{figure1}(a). Due to the large impedance mismatch between air and silicon, acoustic waves are effectively confined within the fully suspended rectangular waveguide. In the design of the anti-resonant supporting arm, we draw inspiration from the suspended anti-resonant acoustic waveguide \cite{SARAW}. By capitalizing on the geometrical softening within the etched slots \cite{SARAW}, we have managed to achieve equivalent anti-resonant reflecting layers to confine the acoustic mode within the central rectangular waveguide. Guided by this principle, the anti-resonant supporting arms not only effectively mitigate the undesired leakage of acoustic modes but also offer structural support for the suspended rectangular waveguide. In Fig. \ref{figure1}(b)-(d), the rectangular waveguide has a width of 1200 nm and a height of 220 nm. The choice of waveguide width 1200 nm aims to achieve large Brillouin amplification, striking a balance between optical loss and photon-phonon overlap. The cross-sectional parameters of the anti-resonant supporting structure (Fig. \ref{figure1}(e)-(g)) refer to Ref. \cite{SARAW}. Optical modes are depicted in Fig. \ref{figure1}(b) and (e), and the corresponding electrostrictive forces are illustrated in Fig. \ref{figure1}(e) and (f). The consistent distributions of the optical modes render the mismatch loss virtually negligible. Fig. \ref{figure1}(d) and (g) display the elastic displacement fields of the forward SBS acoustic modes, exhibiting similar distributions to the electrostrictive force (Fig. \ref{figure1}(e) and (f)), signifying substantial photon-phonon overlap between the acoustic and optical modes. The aforementioned physical field simulations were conducted using the finite element solver COMSOL.

Typically, three-dimensional fabrication of the suspended nanowire requires a series of lithography exposure and plasma etching steps. Yet, in this work, harnessing the loading effect offers a transformative simplification of the entire fabrication process. The loading effect, also known as reactive ion etching (RIE) lag or aspect ratio-dependent etching (ARDE) \cite{LE2}, arises from reactant depletion during the etching process, leading to etching depths determined by the aperture size or local pattern density, as illustrated in Fig. \ref{figure2}(a). This effect is generally perceived as a hindrance in the field of microelectronics because it has the potential to introduce inconsistencies in device design and manufacturing. However, with ingenious design strategies and precise control, we can quantitatively characterize the relationship between slot width and etching depth (Fig. \ref{figure2}(a)-(b)). It empowers us to exploit the loading effect to achieve three-dimensional fabrication. 

As illustrated in Fig. \ref{figure2}(c), our fabrication process begins with the spin coating of the positive electron beam photoresist, followed by EBL exposure of the designed waveguide pattern. After development, the designed pattern is transferred to the silicon-on-insulator (SOI) wafer (crystal orientation, \mbox{$\langle100\rangle$}) through ICP etching. During the ICP etching process, the etched slots within the anti-resonant supporting structure (right dashed box in Fig. \ref{figure2}(c)) exhibit a relatively small width. Consequently, after ICP etching, a certain thickness of silicon film is retained to establish stable connection and support for the central waveguide. On the contrary, the width of the exposed slots in the suspended rectangular waveguide patterned region (left dashed box in Fig. \ref{figure2}(c)) exceeds 1 $\mu m$, leading to over-etching of the silicon layer and exposing the underlying silica layer, which is subsequently eliminated through wet etching utilizing 10\% hydrofluoric acid solution. This loading-effect-based three-dimensional microfabrication approach requires only one lithography exposure and one plasma etching step, significantly reducing manufacturing costs and eliminating overlay exposure. It not only improves the fabrication precision but also reduces the inhomogeneous broadening of the Brillouin resonance introduced by the fabrication process \cite{SARAW}. The SEM images of the fabricated devices are illustrated in Fig. \ref{figure1}(h)-(j), highlighting remarkable structural stability.

\begin{figure}[!t]
\begin{center}
\hspace{0.5cm}
\centering
\includegraphics[width=9cm]{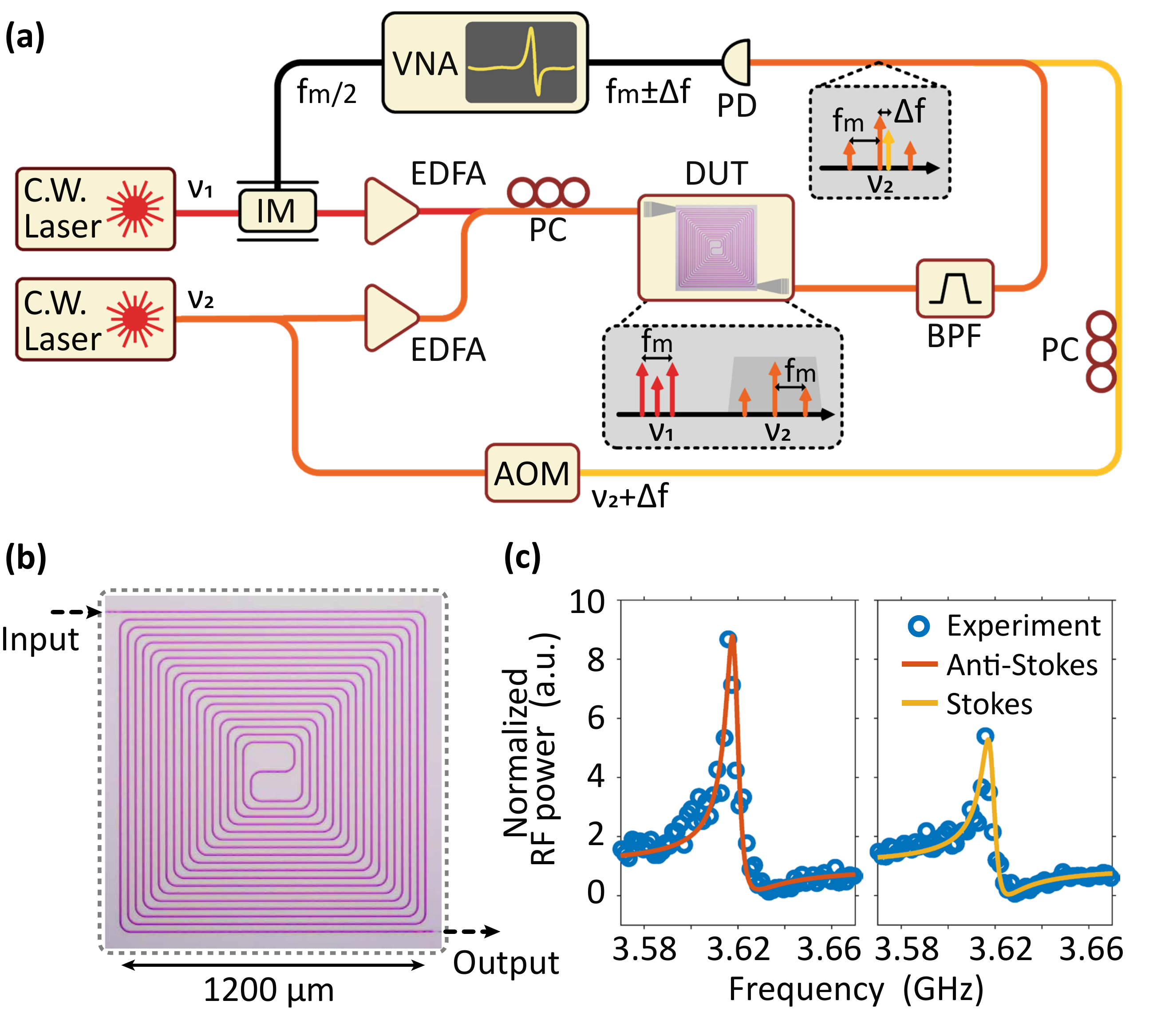}
\caption{(a) The setup of the heterodyne FWM experiment. C.W.: continuous-wave lasers; IM: intensity modulator; VNA: vector network analyzer;PC: polarization controller; EDFA: erbium-doped fiber amplifier; DUT: device under test; AOM: acousto-optic modulator; BPF: band-pass filter; PD: photodetector. (b) Microscopic image of a 5-cm-long suspended nanowire, meticulously shaped into a rectangular spiral with a footprint of 1200 $\mu m \times$ 1200 $\mu m$. (c) Measurement and fitting results of the heterodyne FWM experiment.}
\label{figure3}
\end{center}
\end{figure}

To explore the Brillouin optomechanical characteristics of the suspended nanowire structure, we fabricated a 5-cm-long waveguide and carried out a heterodyne four-wave mixing (FWM) experiment \cite{Mat4,PP2,SARAW}. This measurement scheme is insensitive to fluctuations of input optical power, ensuring robust and reliable results. The experimental configuration, as shown in Fig. \ref{figure3}(a), involves two C.W. lasers operating at frequencies $\nu_1$ and $\nu_2$ with corresponding wavelengths of 1550 and 1552 nm. The upper laser branch is modulated by an intensity modulator with the modulation signal ($f_m/2$) provided by the VNA. After amplified by an EDFA, the modulated sidebands act as the pump light injected into the DUT. At the lower part, the second laser is divided into two paths. The upper path is amplified by an EDFA as the probe light injected into the DUT. The lower path experiences a frequency shift of $\Delta f$ through the AOM to serve as a reference signal. Inside the DUT, the FWM interaction of the pump light ($\nu_1\pm f_m/2$) and the probe light ($\nu_2$) produces two sidebands at $\nu_2\pm f_m$, as illustrated in the gray area beneath the DUT. Followed by a BPF, the FWM-generated sidebands $\nu_2\pm f_m$ are filtered out, and subsequently mixed with the reference signal $\nu_2+\Delta f$ via a PD, generating signals at $f_m\pm\Delta f$. When $f_m$ scans over the Brillouin frequency shift, the intensity variation of sidebands $f_m\pm\Delta f$ are recorded by the VNA, with $f_m+\Delta f$ and $f_m-\Delta f$ corresponding to Stokes and anti-Stokes component respectively. For the 5-cm-long suspended nanowire, we employed a rectangular spiral layout (Fig. \ref{figure3}(b)). This configuration has a compact footprint of 1200 $\mu m \times$ 1200 $\mu m$, effectively mitigating the inhomogeneous broadening introduced by fabrication defects and crystal orientation \cite{SARAW}. By fitting the experimental data (Fig. \ref{figure3}(c)) \cite{Mat4,PP2,SARAW}, we obtained the G$_B$ of this spiral waveguide is 1100 W$^{-1}$m$^{-1}$, with a Brillouin frequency shift of 3.62 GHz and a Q$_m$ of 550. These results align with our simulations, which predicted a G$_B$ of 1150 W$^{-1}$m$^{-1}$ and an acoustic mode eigenfrequency of 3.69 GHz. To the best of our knowledge, this is the highest Brillouin gain coefficient in silicon waveguides with comparable length and cross-sectional dimensions. It underscores that the proposed suspended nanowire structure can support an exceptionally strong on-chip Brillouin optomechanical interaction with a remarkably simple design and fabrication process.

\begin{figure}[!t]
\begin{center}
\hspace{0.5cm}
\centering
\includegraphics[width=9cm]{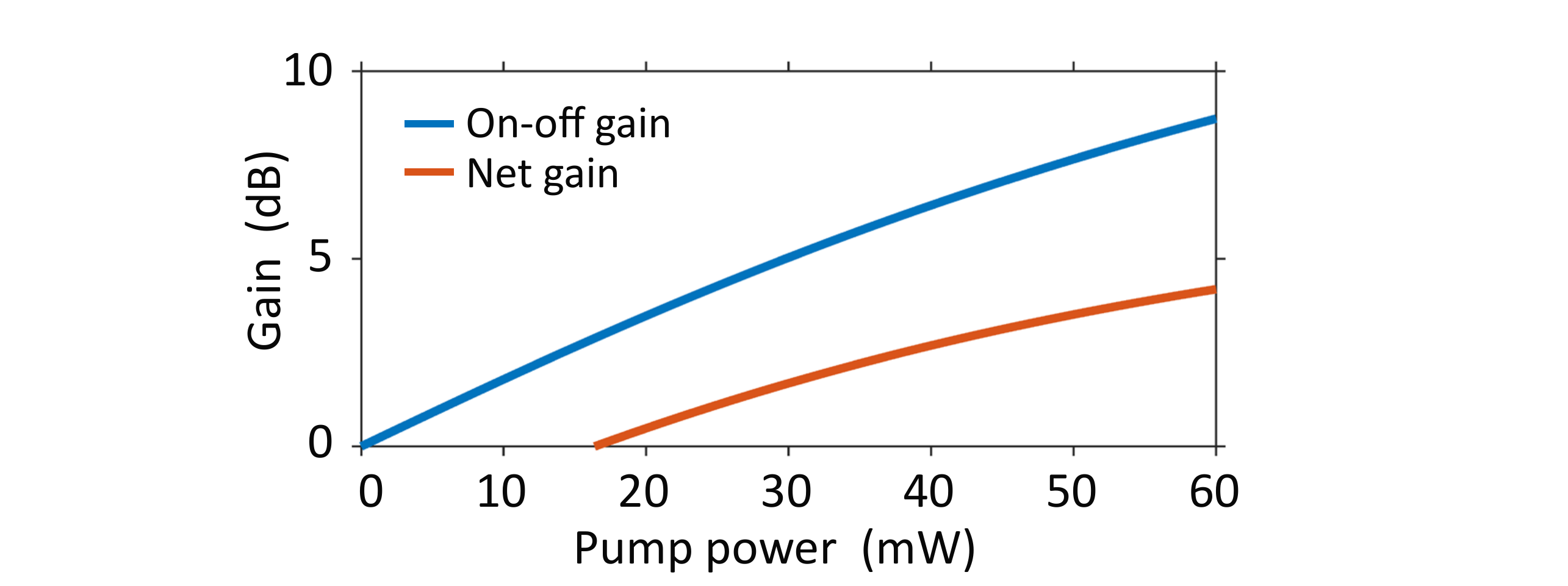}
\caption{Numerically calculated Brillouin on-off gain and net gain of the 5-cm-long suspended nanowire.}
\label{figure4}
\end{center}
\end{figure}

Our fabricated device exhibits a linear optical loss of 0.5 dB/cm. Considering the nonlinear absorption losses \cite{SARAW}, we numerically calculated the Brillouin on-off gain and net gain for a 5-cm-long suspended nanowire with the G$_B$ of 1100 W$^{-1}$m$^{-1}$ (Fig. \ref{figure4}). With an on-chip pump power of 60 mW, we can achieve an on-off gain of 8.7 dB and a net gain of 4.1 dB. This marks a state-of-the-art level of on-chip Brillouin amplification within silicon waveguides. Nonetheless, the current limitations of these results are primarily attributed to the waveguide sidewall roughness, which restricts the further optimization of optical losses and Q$_m$. Referring to prior research \cite{Mat4}, the integration of a CMOS pilot line in the foundry has the potential to yield reduced optical losses and enhanced $Q_m$. These advancements hold the promise of achieving larger Brillouin gain.

In conclusion, we present for the first time a on-chip suspended nanowire structure utilizing loading-effect-based three-dimensional microfabrication. This structure offers extremely streamlined design and fabrication processes, involving a single-step lithography exposure. Utilizing a compact spiral waveguide layout, we have achieved a G$_B$ of 1100 W$^{-1}$m$^{-1}$ in a 5-cm-long waveguide with a footprint of merely 1200 $\mu m \times$ 1200 $\mu m$. Due to the ubiquitous presence of the loading effect in both dry and wet etching processes, this waveguide structure and fabrication method are universal and can be seamlessly transplanted to other integrated photonic platforms, such as silicon nitride and chalcogenide glass. Additionally, even if the geometric parameters at the anti-resonant supporting arm deviate from the design due to fabrication inaccuracies, the Brillouin gain coefficient remains largely unaffected. This resilience stems from the small proportion of the length of the supporting arms. As the majority of the nanowire remains fully suspended, effective confinement of the acoustic modes for forward SBS is sustained. Notably, our scheme eliminates the need for overlay lithography exposure, minimizing the performance requirements of the fabrication equipment. In light of these advantages, our proposed structure and fabrication scheme for on-chip Brillouin optomechanics present a significantly elevated fabrication convenience, a superior yield rate, and an enhanced photon-phonon coupling intensity. It is well-suited for large-scale integration of Brillouin-based photonic devices, opening up new avenues for three-dimensional microfabrication in on-chip acousto-optic interaction platforms.

\bibliography{mybib}

\begin{thebibliography}{10}
\newcommand{\enquote}[1]{``#1''}

\bibitem{Review1}
M.~Merklein, I.~V. Kabakova, A.~Zarifi, and B.~J. Eggleton, \enquote{\href{https://doi.org/10.1063/5.0095488}{100 years of Brillouin scattering: Historical and future perspectives},} {\protect\JournalTitle{Applied Physics Reviews}} \textbf{9}, 041306 (2022).

\bibitem{Review2}
B.~J. Eggleton, C.~G. Poulton, P.~T. Rakich, M.~J. Steel, and G.~Bahl, \enquote{\href{https://doi.org/10.1038/s41566-019-0498-z}{Brillouin integrated photonics},} {\protect\JournalTitle{Nature Photonics}} \textbf{13}, 664--677 (2019).

\bibitem{Filter2}
M.~Garrett, Y.~Liu, M.~Merklein, C.~T. Bui, C.~K. Lai, D.-Y. Choi, S.~J. Madden, A.~Casas-Bedoya, and B.~J. Eggleton, \enquote{\href{https://doi.org/10.1038/s41467-023-43404-x}{Integrated microwave photonic notch filter using a heterogeneously integrated Brillouin and active-silicon photonic circuit},} {\protect\JournalTitle{Nature Communications}} \textbf{14}, 7544 (2023).

\bibitem{IMP5}
S.~Gertler, N.~T. Otterstrom, M.~Gehl, A.~L. Starbuck, C.~M. Dallo, A.~T. Pomerene, D.~C. Trotter, A.~L. Lentine, and P.~T. Rakich, \enquote{\href{https://doi.org/10.1038/s41467-022-29590-0}{Narrowband microwave-photonic notch filters using Brillouin-based signal transduction in silicon},} {\protect\JournalTitle{Nature Communications}} \textbf{13}, 1947 (2022).

\bibitem{Laser4}
W.~Loh, J.~Stuart, D.~Reens, C.~D. Bruzewicz, D.~Braje, J.~Chiaverini, P.~W. Juodawlkis, J.~M. Sage, and R.~McConnell, \enquote{\href{https://doi.org/10.1038/s41586-020-2981-6}{Operation of an optical atomic clock with a Brillouin laser subsystem},} {\protect\JournalTitle{Nature}} \textbf{588}, 244--249 (2020).

\bibitem{IOSP3}
S.~Gundavarapu, G.~M. Brodnik, M.~Puckett, T.~Huffman, D.~Bose, R.~Behunin, J.~Wu, T.~Qiu, C.~Pinho, N.~Chauhan, J.~Nohava, P.~T. Rakich, K.~D. Nelson, M.~Salit, and D.~J. Blumenthal, \enquote{\href{https://doi.org/10.1038/s41566-018-0313-2}{Sub-hertz fundamental linewidth photonic integrated Brillouin laser},} {\protect\JournalTitle{Nature Photonics}} \textbf{13}, 60--67 (2019).

\bibitem{Sensor1}
C.~Galindez-Jamioy, J.~López-Higuera, and R.~Bernini, \enquote{\href{https://doi.org/10.1155/2012/204121}{Brillouin Distributed Fiber Sensors: An Overview and Applications},} {\protect\JournalTitle{Journal of Sensors}} \textbf{2012}, 204121 (2012).

\bibitem{Sensor3}
X.~Bao and L.~Chen, \enquote{\href{https://www.mdpi.com/1424-8220/11/4/4152}{Recent Progress in Brillouin Scattering Based Fiber Sensors},} {\protect\JournalTitle{Sensors}} \textbf{11}, 4152--4187 (2011).

\bibitem{cool}
W.~Renninger, P.~Kharel, R.~Behunin, and P.~Rakich, \enquote{\href{https://doi.org/10.1038/s41567-018-0090-3}{Bulk crystalline optomechanics},} {\protect\JournalTitle{Nature Physics}} \textbf{14}, 601--607 (2018).

\bibitem{IOSP2}
E.~A. Kittlaus, N.~T. Otterstrom, P.~Kharel, S.~Gertler, and P.~T. Rakich, \enquote{\href{https://doi.org/10.1038/s41566-018-0254-9}{Non-reciprocal interband Brillouin modulation},} {\protect\JournalTitle{Nature Photonics}} \textbf{12}, 613--619 (2018).

\bibitem{Exc}
Y.-H. Lai, Y.-K. Lu, M.-G. Suh, Z.~Yuan, and K.~Vahala, \enquote{\href{https://doi.org/10.1038/s41586-019-1777-z}{Observation of the exceptional-point-enhanced Sagnac effect},} {\protect\JournalTitle{Nature}} \textbf{576}, 65--69 (2019).

\bibitem{Mat1}
B.~Morrison, A.~Casas-Bedoya, G.~Ren, K.~Vu, Y.~Liu, A.~Zarifi, T.~G. Nguyen, D.-Y. Choi, D.~Marpaung, S.~J. Madden, A.~Mitchell, and B.~J. Eggleton, \enquote{\href{https://opg.optica.org/optica/abstract.cfm?URI=optica-4-8-847}{Compact Brillouin devices through hybrid integration on silicon},} {\protect\JournalTitle{Optica}} \textbf{4}, 847--854 (2017).

\bibitem{Mat3}
R.~Van~Laer, B.~Kuyken, D.~Van~Thourhout, and R.~Baets, \enquote{\href{https://doi.org/10.1038/nphoton.2015.11}{Interaction between light and highly confined hypersound in a silicon photonic nanowire},} {\protect\JournalTitle{Nature Photonics}} \textbf{9}, 199--203 (2015).

\bibitem{Mat4}
E.~A. Kittlaus, H.~Shin, and P.~T. Rakich, \enquote{\href{https://doi.org/10.1038/nphoton.2016.112}{Large Brillouin amplification in silicon},} {\protect\JournalTitle{Nature Photonics}} \textbf{10}, 463--467 (2016).

\bibitem{Mat9}
R.~Botter, K.~Ye, Y.~Klaver, R.~Suryadharma, O.~Daulay, G.~Liu, J.~van~den Hoogen, L.~Kanger, P.~van~der Slot, E.~Klein, M.~Hoekman, C.~Roeloffzen, Y.~Liu, and D.~Marpaung, \enquote{\href{https://www.science.org/doi/abs/10.1126/sciadv.abq2196}{Guided-acoustic stimulated Brillouin scattering in silicon nitride photonic circuits},} {\protect\JournalTitle{Science Advances}} \textbf{8}, eabq2196 (2022).

\bibitem{Mat5}
D.~B. Sohn, S.~Kim, and G.~Bahl, \enquote{\href{https://doi.org/10.1038/s41566-017-0075-2}{Time-reversal symmetry breaking with acoustic pumping of nanophotonic circuits},} {\protect\JournalTitle{Nature Photonics}} \textbf{12}, 91--97 (2018).

\bibitem{PP2}
K.~Wang, M.~Cheng, H.~Shi, L.~Yu, C.~Huang, S.~Qin, Y.~Zhang, L.~Kai, and J.~Sun, \enquote{\href{https://doi.org/10.1021/acsphotonics.1c00880}{Demonstration of Forward Brillouin Gain in a Hybrid Photonic–Phononic Silicon Waveguide},} {\protect\JournalTitle{ACS Photonics}} \textbf{8}, 2755--2763 (2021).

\bibitem{Structure1}
P.~N. Ruano, J.~Zhang, D.~Melati, D.~Gonz{\'a}lez-Andrade, X.~Le~Roux, E.~Cassan, D.~Marris-Morini, L.~Vivien, N.~D. Lanzillotti-Kimura, and C.~Alonso-Ramos, \enquote{\href{https://doi.org/10.1016/j.optlastec.2023.109130}{Genetic optimization of Brillouin scattering gain in subwavelength-structured silicon membrane waveguides},} {\protect\JournalTitle{Optics \& Laser Technology}} \textbf{161}, 109130 (2023).

\bibitem{SARAW}
P.~Lei, M.~Xu, Y.~Bai, Z.~Chen, and X.~Xie, \enquote{\href{https://arxiv.org/abs/2401.12677v1}{Anti-resonant acoustic waveguides enabled tailorable Brillouin scattering on chip},}  (2024).

\bibitem{LE2}
K.~Gantz, L.~Renaghan, and M.~Agah, \enquote{\href{https://dx.doi.org/10.1088/0960-1317/18/2/025003}{Development of a comprehensive model for RIE-lag-based three-dimensional microchannel fabrication},} {\protect\JournalTitle{Journal of Micromechanics and Microengineering}} \textbf{18}, 025003 (2007).

\end{thebibliography}

\bibliographyfullrefs{mybib}
\end{document}